# Explainable Sustainability for AI in the Arts

PETRA JÄÄSKELÄINEN, KTH Royal Institute of Technology, Sweden

AI is becoming increasingly popular in artistic practices, but the tools for informing practitioners about the environmental impact (and other sustainability implications) of AI are adapted for other contexts than creative practices - making the tools and sustainability implications of AI not accessible for artists and creative practitioners. In this position paper, I describe two empirical studies that aim to develop environmental sustainability reflection systems for AI Arts, and discuss and introduce Explainable Sustainability in for AI Arts.

## 1 INTRODUCTION

The recent developments of artificial intelligence increase its capabilities for artistic practices. However, the use of AI technology comes bearing an environmental cost that has been criticized in general AI research. There have been efforts to quantify [3, 4, 8, 9] and steer development of AI [12, 13] towards more sustainable approaches (in most cases this means reduced energy expenditure) - but in the domain of AI for the arts sustainability has been largely ignored [5–7]. As [6] describe, AI for the Arts could potentially have high environmental costs depending on the scale and use cases of the adoption of AI in artistic contexts. For example, individual contemporary arts projects may not accumulate a significant impact, but, for example, large-scale adoption in media generation (imagine future AI generated Netflix[1] content or Metaverse environments[2]), the alarming scale of the potential environmental impact starts to reveal itself.

Thus, [6] have proposed the research agenda of environmental sustainability of AI Arts and highlighted the problems within that. These problems, for example, include drawing boundaries to where these practices start and end (if artists are mixing various materials and approaches), as well as the difficulties in quantifying environmental impact, and also the limitations that a quantifying approach might bear (quantification can not encompass all that there is to sustainability, that is). Building on this prior research and acknowledging these challenges, the take on this position paper of the environmental impact of AI Art proposes *Explainable Sustainability for AI Arts* - efforts towards developing methods and tools for informing the users about the environmental impact of their practices - and building systems that are transparent of their sustainability impact, with the potential end goal of facilitating change towards more sustainable practices in AI Arts. I would particularly like to point out, that 1) these methods and tools do not need to consist of technological solutions, but may utilize a variety of methods in achieving these goal (as technological solutionism is one of the large problems from sustainability point-of-view) and 2) that the complexity of climate crisis and sustainability phenomena can not only be encompassed by explainability regarding energy consumption, CO2, or other quantified measures that relate to the immediate use of the system. However, these quantifications can provide a feasible starting point for practical research explorations in Explainable Sustainability for AI arts. As two cases of our on-going work, I present 1) The Green Notebook and 2) Visual Feedback System for Environmental Impact as two exploratory and preliminary cases of explainable environmental sustainability for AI in the arts.

---

[1] https://www.iea.org/commentaries/the-carbon-footprint-of-streaming-video-fact-checking-the-headlines
[2] https://www.weforum.org/agenda/2022/02/how-metaverse-actually-impacts-the-environment/

## 2 SUSTAINABILITY OF AI ART AND EXPLAINABLE SUSTAINABILITY FOR AI ARTS

While there is limited amount of prior research on sustainability of AI Art, one of the central papers regarding it [6] has proposed research agenda in sustainability and AI Arts, while also providing a conceptual framework that takes into account different stages of the artists' process and different AI hardware used in these stages. It is also highlighted in the paper how this type of quantified analysis has its limitations and in absence of political steering (or to add - in aims for facilitating cultural change) the environmental impact of AI arts is likely to be a concern [6]. In this paper, I approach environmental sustainability specifically from the perspective of facilitating cultural change, and build on an assumption that in order to create such change, the users of these systems should firstly understand what the environmental impact of their actions might be. This is, where *Explainable Sustainability for AI in the Arts* comes in.

As [1] note, there is a significant lack of explainable AI research for the arts. While explainable AI is well established approach for designing AI systems that is aiming for providing good and relevant explanations about the reasonings of the AI system to the user, when it comes to Explainable Sustainability - and more specifically environmental sustainability - relevant questions arise regarding what explanations are good for users in different contexts? We propose this as one of the central questions for *Explainable Sustainability for AI in the Arts*. Environmental sustainability in some cases can be approaches as CO2 emissions or energy impact, but sustainability as a phenomena is more complex than that. It involves relation to planetary boundaries [11], users attitudes towards sustainability - as a few examples that are harder to explain in and by the system. Nevertheless, we propose that explainable sustainability for AI Arts should entail the immediate measures of the AI Art technology itself (energy consumption, CO2, life cycle) but also aim to inform the users about these more abstract concepts that relate to sustainability.

## 3 CASE 1: THE GREEN NOTEBOOK

The Green Notebook [2] is a design artifact that was developed in an exploratory RtD study to explore notebook-based environmental sustainability reflection. Essentially, the diary resembled a workbook/notebook where artists could write down their ideas and the Notebook would provide them information about the environmental sustainability of their choices (energy consumption that relates to choices of GPU, hardware, AI training, etc.) and prompt questions that would facilitate guided sustainability reflection (see Fig. 1 for image of the prototype). In this study, many insights emerged (which we can not cover sufficiently in the scope of this paper but rather highlight the key findings). One of them being varying conversational strategies used by artists. In [2], we propose two dimensions for conversational strategies that artists used when engaging in such sustainability reflection: command vs. conversation based and abstract vs. specific. These identified strategies can be taken into consideration when developing conversational explainable sustainability systems for AI arts. A particular challenge in artists' communication about their work was translating the abstract and conversation-based commands into concrete information about the environmental impact. Furthermore, design trade-offs found in the context of the Notebook study entailed efficiency vs. politeness, and focus vs. integration. The efficient conversational strategies were more information based, whereas polite conversation strategies involved more filler words, and artists seemed to have differing preferences regarding them. The Notebook itself was prioritizing focus (being de-attached from the artistic process), while in our other prototype in Case 2 (see Section 4) an integrated tool design was explored.

## 4 CASE 2: INTEGRATED ENVIRONMENTAL IMPACT FEEDBACK SYSTEM

In case 2 [10], we explored a system design that integrated environmental sustainability to the system interface itself. Thus, this design approach was an integrative approach in contrast to the Case 1 (see Section 3, focus vs. integration). This prototype included color coded and graphic feedback of the energy consumption of one specific browser-based generative AI system (see Fig. 1). This system changed the visual feedback based on choices that the user made, so that the environmental sustainability feedback would be presented in real-time to the users. With increasing energy consumption, the prototype changed into red, while intermediate consumption was represented by orange, and a light consumption with green. The visual feedback was informed by a survey study of associations regarding energy consumption and environmental sustainability, including colors, symbols, and graphs. Such feedback can prove effective, but as our initial results suggest, caution should be practiced in utilizing such feedback. For example, providing green feedback might contribute into green-washing by giving the impression that users are working in a sustainable manner, while the overarching consumption of all users across the world may still have a significant environmental impact. Thus, these design strategies may be useful in prompting individual users about their consumption in specific tasks, but designers of the systems need to be mindful of their definition of the consumption and the notion of sustainability in a wider context.

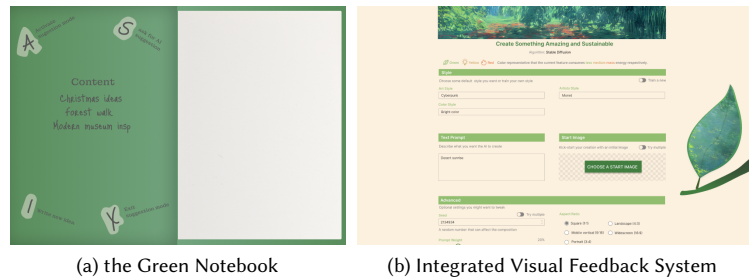

(a) the Green Notebook        (b) Integrated Visual Feedback System

Fig. 1. Case 1 and Case 2: Two different types of sustainability reflection systems, prioritizing focus vs. integration

## 5 DISCUSSION AND CONCLUSION

In this position paper, we have introduced two RtD studies on explainable environmental sustainability, with a focus on informing AI artists about energy consumption of their practices. We have also discussed dimensions that would be relevant to include in efforts towards explainable sustainability in the future that go past the immediate energy usage of the system. We suggest, that while simultaneously exploring the design space and concrete strategies for changing user behavior towards reduction of computing resources and Co2, these systems should also aim to explain matters that are difficult to quantify but nevertheless contribute in the state of sustainability. For example, [7] has explored earlier cases in which users of AI Art systems are not aware of the complex structures regarding the environmental exploitation that lie behind the technologies (such as mineral mining and capitalist related exploitation of labourers in low wage countries) and suggested exploring how these distanced aspects could be brought closer to the users of the technologies. We anticipate, that the core challenge of Explainable Sustainability of AI arts will lie in how these complex causalities, impacts, and processes will and can be brought to the users of the reductionist systems in a manner that will facilitate knowledge-building, behavior change, and shift in attitudes and values towards sustainability.